# Deep sub-micron stud-via technology for superconductor VLSI circuits

**Sergey K. Tolpygo, V. Bolkhovsky, T. Weir, L.M. Johnson, W.D. Oliver, and M.A. Gouker**

Lincoln Laboratory, Massachusetts Institute of Technology, Lexington, MA 02420, USA

Sergey.Tolpygo@ll.mit.edu

**Abstract**. A fabrication process has been developed for fully planarized Nb-based superconducting inter-layer connections (vias) with minimum size down to 250 nm for superconductor very large scale integrated (VLSI) circuits with 8 and 10 superconducting layers on 200-mm wafers. Instead of single Nb wiring layers, it utilizes Nb/Al/Nb trilayers for each wiring layer to form Nb pillars (studs) providing vertical connections between the wires etched in the bottom layer of the trilayer and the next wiring layer that is also deposited as a Nb/Al/Nb trilayer. This technology makes possible a dramatic increase in the density of superconducting digital circuits by reducing the area of interconnects with respect to presently utilized etched contact holes between superconducting layers and by enabling the use of stacked vias. Results on the fabrication and size dependence of electric properties of Nb studs with dimensions near the resolution limit of 248-nm photolithography are presented in the normal and superconducting states. Superconducting critical current density in the fabricated stud-vias is about 0.3 A/$\mu m^2$ and approaches the depairing current density of Nb films.

## 1. Introduction

In order to realize the tremendous advantages of superconducting digital integrated circuits over semiconductor circuits in speed and reduction of energy dissipation [1], their integration scale must be increased from its current medium level to very large scale integration (VLSI) levels and beyond. This requires increasing the density of elemental switching devices – Josephson junctions (JJs) – by several orders of magnitude, from the present density of ~ $10^4$ - $10^5$ JJs per $cm^2$ to $10^6$ - $10^7$ JJs/$cm^2$, with a corresponding increase in the density of interconnects (Josephson transmission and passive transmission lines) and other passive components (inductors, resistors, vias). These goals can be achieved by scaling down the dimensions of all circuit elements and increasing the number of superconducting metal layers available for circuit integration, a path familiar from the development of semiconductor VLSI and ULSI circuits. Currently, the most advanced fabrication process for Nb-based superconducting circuits utilizes up to 10 superconducting metal (Nb) layers, chemical-mechanical polishing (CMP) for planarization of dielectric and Nb layers below the junctions layer, minimum JJ size of 1 µm, minimum linewidth ~0.5-µm, and reaches JJ density ~ $10^5$ JJs per chip [2].

Increasing the density of Josephson junctions and the number of superconducting layers leads to an even larger increase in the number and density of vias between metal layers. Although scaling down the dimensions of Josephson junctions, inductors, and resistors is more or less straightforward by implementing modern deep-UV photolithography and etch tools, the reduction of the sizes of vias between superconducting layers is not. In all existing and reported fabrication processes [2]-[10], superconducting vias are made by etching contact holes in the inter-metal dielectric and depositing the next Nb layer by physical vapor deposition (PVD), usually by dc magnetron sputtering, as shown schematically in figure 1. This limits the minimum size of contact holes to ~ 0.5 µm, as it becomes difficult to achieve reliable contacts with sufficient superconducting critical currents in smaller holes, because increasing their aspect ratio leads to poor step coverage and formation of voids (keyholes). Sloping the contact walls by usual methods to improve step coverage becomes more difficult with the transition to deep-UV lithography which favors near-vertical photoresist profiles.

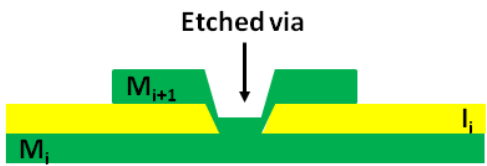

Figure 1. Etched via between metals $M_i$ and $M_{i+1}$. The via is obtained by etching a sloped contact hole in interlayer dielectric $I_i$ that is then filled with sputtered metal $M_{i+1}$. Contact holes with diameters less than ~ 500 nm are difficult to fill by sputtered Nb.

In practice, the contact hole also needs to be well surrounded by metal overlay to allow for photolithography misalignment (e.g., overlay mean plus three sigma) and etch bias. In the typical design rules, this surround is also ~ 0.5 µm making the typical smallest via size around 1 µm. This size further increases when many vias need to be stacked on top of each other to make low-inductance connections between bottom and top layers in a 10-metal layer circuit. As a result, vias became one of the largest componenets in superconducting digital circuits, occupying up to 30% of the circuit area because the number of vias is much larger than the number of Josephson junctions. Clearly, increasing the integration scale of superconducting circuits requires scaling down the size of vias without compromising their reliability or lowering their critical currents.

In the semiconductor industry, a similar problem with contact holes was solved a long ago by developing tungsten plugs between wiring layers [11]-[12], a process that replaced aluminum PVD with W chemical vapor deposition (CVD) to fill high-aspect-ratio contact holes, and metal CMP to remove W from horizontal surfaces of the inter-metal dielectric. Although a similar damascene-type process is possible for Nb circuits, to the best of our knowledge, there is neither a Nb CVD (or PVD) nor a Nb CMP processes available that are capable of producing superconducting Nb films and plugs, and their development for VLSI circuits would be very expensive judging by the experience of the semiconductor industry with the tungsten plug development.

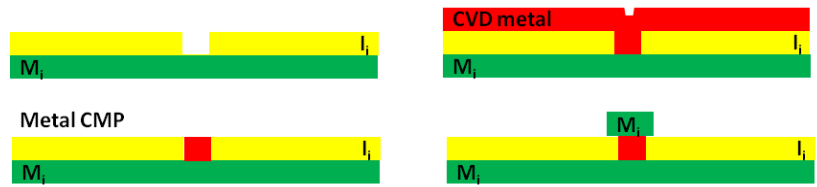

Figure 2. Schematic process of a plug via between metals $M_i$ and $M_{i+1}$ (from left to right and top to bottom): etching of a contact hole with near-vertical walls in $SiO_2$ interlayer dielectric $I_i$, conformal deposition of a plug metal (e.g., W in CMOS circuits) by CVD, CMP of the deposited metal to form a plug, deposition and etching of the wiring metal $M_{i+1}$. A similar process for Nb circuits would require CVD and CMP of Nb, or some other superconducting metal or alloy fully compatible with Nb, without compromising superconducting properties of Nb and the plug material.

In this paper we propose and describe a new process for deep-submicron vias that uses etched Nb studs between wiring layers instead of etched contact holes with metal filling. The studs are formed by Nb etching with subsequent dielectric deposition and dielectric CMP processes. We present results of the practical realization of this process, including electrical characterization of the resultant vias at room and LHe temperatures. We found that this process allowed us to form Nb stud-vias with sizes down to our photolithography resolution limit (~250 nm) and with superconducting critical current density approaching the depairing critical current density of Nb.

## 2. Stud-via process description

The problems associated with metal filling of small-diameter contact holes in interlayer dielectric can be easily avoided if we instead etch cylindrically shaped Nb studs (with nearly vertical walls) from Nb film in contact with bottom wiring layer $M_i$. Then, an interlayer dielectric could be deposited and planarized using CMP to make access to the top of the stud. After this the structure would look exactly as shown in the left bottom picture in figure 2. Then, the top wiring layer deposition would follow.

There are two possible practical realization of such a process. The first version, shown in figure 3, is referred to as a single etch and planarization (SEAP) process.

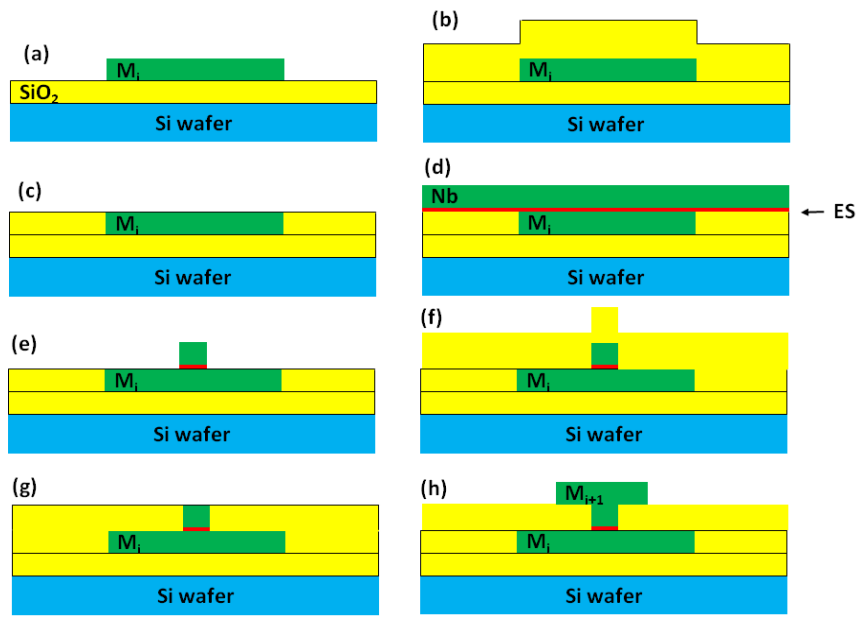

Figure 3. Cross sections of the SEAP version of a stud-via process for superconducting circuits: (a) – deposition and pattering of (Nb) wiring layer $M_i$ over the wafer; (b) – deposition of a blanket $SiO_2$ layer for CMP planarization; (c) CMP of the $SiO_2$ layer; (d) deposition of an etch-stop layer (ES) and Nb stud metal bi-layer; (e) patterning of Nb studs and ES layer; (f) blanket $SiO_2$ deposition for CMP planarization; (g) CMP of $SiO_2$ to access the top of Nb studs; (h) deposition and pattering of the next Nb wiring layer, $M_{i+1}$. Prior step (a) the wafer may already contain several patterned and planarized layers. A convenient etch stop layer is Al. The bi-layer deposition should be done in-situ to minimize contamination of the interface between Al and Nb stud.

Here, the bottom Nb wiring layer $M_i$ is first deposited and patterned, figure 3(a). This layer is then planarized by depositing a thick layer of $SiO_2$ [figure 3(b)] with subsequent CMP to reach the structure shown in figure 3(c). After this, Nb deposition for subsequent studs follows, figure 3(d). Since Nb studs need to be etched while stopping on the bottom Nb wiring, it is useful to have a good etch stop (ES) layer – a material with a lower etch rate than Nb. This ES layer must not degrade the superconducting properties of the interface in order to maintain high critical current in the superconducting state. There are very few materials suitable for this, and a thin layer of aluminium (a few nanometers thick) is one of them. After etching the studs, figure 3(e), the process flow repeats planarization steps (b) and (c) to planarize studs and open access to their top surface, figure 3(g). Finally, the top Nb wiring layer is deposited and patterned to complete the via-stud interconnects between a pair of Nb layers $M_i$ and $M_{i+1}$.

The process can further be repeated for the next pair of wiring layers $M_{i+1}$ and $M_{i+2}$ and so on in a straightforward manner. It also allows for stacking studs on top of each other. The drawback of this processing scheme is the need for two planarization steps to form each stud-via interconnect: one for the bottom wire and another one for the etched stud.

The process can be simplified by noticing that the structure formed after stud etching, figure 3(e), resembles the standard trilayer-type SNS Josephson junction where the base electrode is bottom wiring layer $M_i$, the counter (top) electrode is the stud, and the Al etch stop layer is N-layer or Al-$AlO_x$ barrier layer in tunnel junction trilayers [13] without oxidation. Therefore, the stud-via formation can be done in the same manner as the formation of Josephson tunnel junctions in a whole-wafer trilayer process [13] with the addition of planarization [14]-[17]. This second version of the stud-via process, shown in figure 4, is referred to as a dual-etch and planarization (DEAP) process.

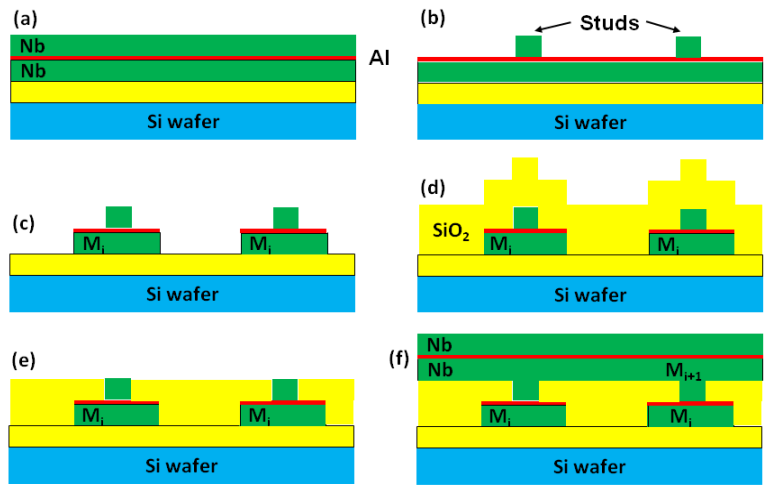

Figure 4. Cross-sections of Nb/Al/Nb trilayer-based (DEAP) stud-via process: (a) Nb/Al/Nb in-situ trilayer deposition; (b) patterning of the top electrode to etch Nb studs; (c) pattering of the bottom electrode to form wiring layer $M_i$; (d) blanket $SiO_2$ deposition for planarization by CMP; (e) CMP down to the tops of Nb studs; (f) Nb/Al/Nb wiring layer deposition to form wiring layer $M_{i+1}$ and Nb stud to contact next layer $M_{i+2}$. If no additional stud-vias are required above, a single Nb wiring layer $M_{i+1}$ can be deposited instead. This process is to be repeated as many times as there are wiring layers in the entire process.

Instead of depositing a single wiring layer as in the previous (SEAP) version, in this process each wiring layer is deposited in-situ as a Nb/Al/Nb trilayer, figure 4(a). The top layer of the trilayer is then patterned by photolithography and dry etching to form Nb studs as shown in figure 4(b). Aluminium serves as an excellent etch-stop layer for etching Nb in F-based chemistries. Then, the bottom electrode of the trilayer is patterned in Cl-based chemistry to etch through the Al layer and form the wiring pattern $M_i$. Etching is followed by blanket $SiO_2$ deposition and CMP to access the tops of Nb studs, figure 3(e). Then, the next layer $M_{i+1}$ is deposited as a Nb/Al/Nb trilayer and patterned in the same manner to form the next layer of interconnects, and so on. If no more stud-vias are required above, a single metal wiring layer can be deposited. The described processing unit needs to be repeated as many times as there are wiring layers in the full process. The only disadvantage of the DEAP process with respect to the SEAP version described earlier is the need to planarize by dielectric CMP twice higher steps in the etched metals (stud plus bottom wire).

Our near-term goal is the development of a 10-metal layer process for superconducting VLSI circuits on 200-mm wafers with deep submicron features with the described stacked stud-vias. The cross-section of this target process is shown in figure 5. The target critical current density of this process is 100 $\mu A/\mu m^2$ (10 $kA/cm^2$), the minimum JJ size is 0.5 µm, the wiring pitch (line + space) is 0.5 µm, and the minimum diameter of Nb stud-vias is 0.3 µm. As a step towards this goal, in the next section we will describe the results of stud-vias fabrication by DEAP version of the process (trilayer-based, figure 4).

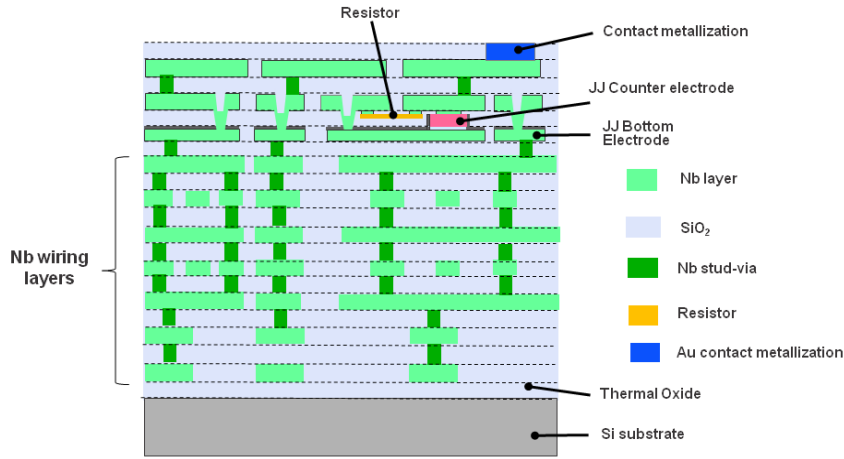

Figure 5. Cross-section of the target process with 10 planarized metal layers and stud-vias for superconducting VLSI circuits. Conventional etched vias are only used to connect to Nb bottom electrode of JJs. Making stud-vias to this layer would be too complicated, because the counter electrodes of JJs are already Nb studs connecting to the upper wiring layer.

### 3. Stud-via fabrication

Nb/Al/Nb trilayers were deposited over 200-mm Si wafers with 500 nm of thermal oxide. The target thickness of the top Nb layer (Nb studs) was 250 nm, and of the bottom layer and the wiring layer was 200 nm. The thickness of the Al etch-stop layer was 8 nm.

*3.1 Photolithography*

The photolithography was done using a positive deep-UV photoresist, bottom antireflection coating, Canon FPA-3000 EX4 248-nm stepper with 5x reduction, and 250-nm nominal resolution. The circular shape for stud definition was implemented with diameters covering the range from 200 nm to 15 µm. Clear field masks were used without implementing any resolution enhancement techniques. The average diameter of the developed photoresist features on wafers was measured using a Hitachi CD SEM, and the typical SEM images are shown in figure 6. The wafers were exposed using a 7 x 7 grid with 22 mm die size, each die containing 16 test chips. Die locations are referred to as c#r# indicating the column and row numbers of this grid, respectively.

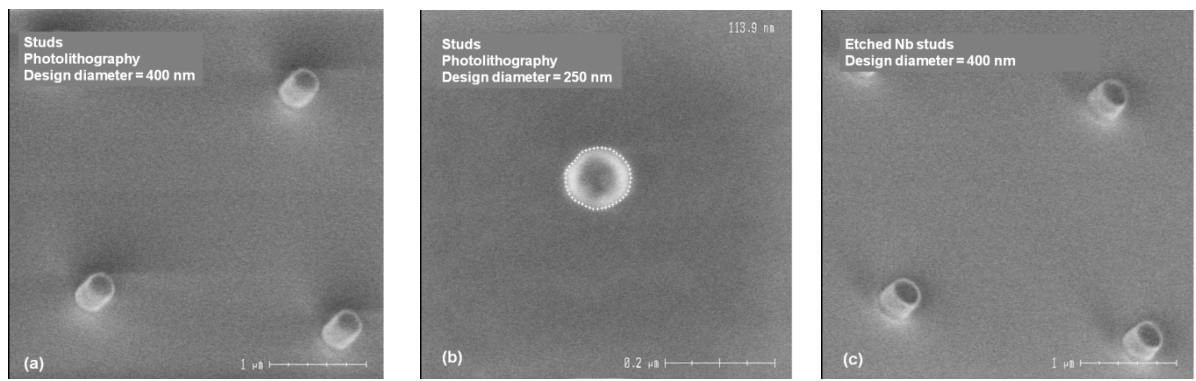

Figure 6. SEM images of the photoresist mask and etched Nb features: (a) tilted view of 400-nm photoresist posts masking Nb/Al/Nb trilayer used for stud-via definition; (b) top view of a photoresist post with design diameter of 250 nm, showing the automated CD measurement of the mean diameter using 48 points around the feature perimeter and giving 113.9 nm (right upper corner); (c) tilted view of Nb studs obtained after etching the top electrode of Nb/Al/Nb trilayer.

Studs photolithography is identical to photolithography of JJs and contact holes (the latter would differ only by the use of dark field masks). We studied carefully the size dependence of the obtained features on wafers, $d_w$, on the size of the design features $d$ (on the reticles the design size is $5d$ to account for 5x reduction of the stepper optics) because these results are applicable to the definition of circular JJs, contact holes, and other features with sizes near the diffraction-limited resolution minimum. The obtained dependences are shown in figure 7.

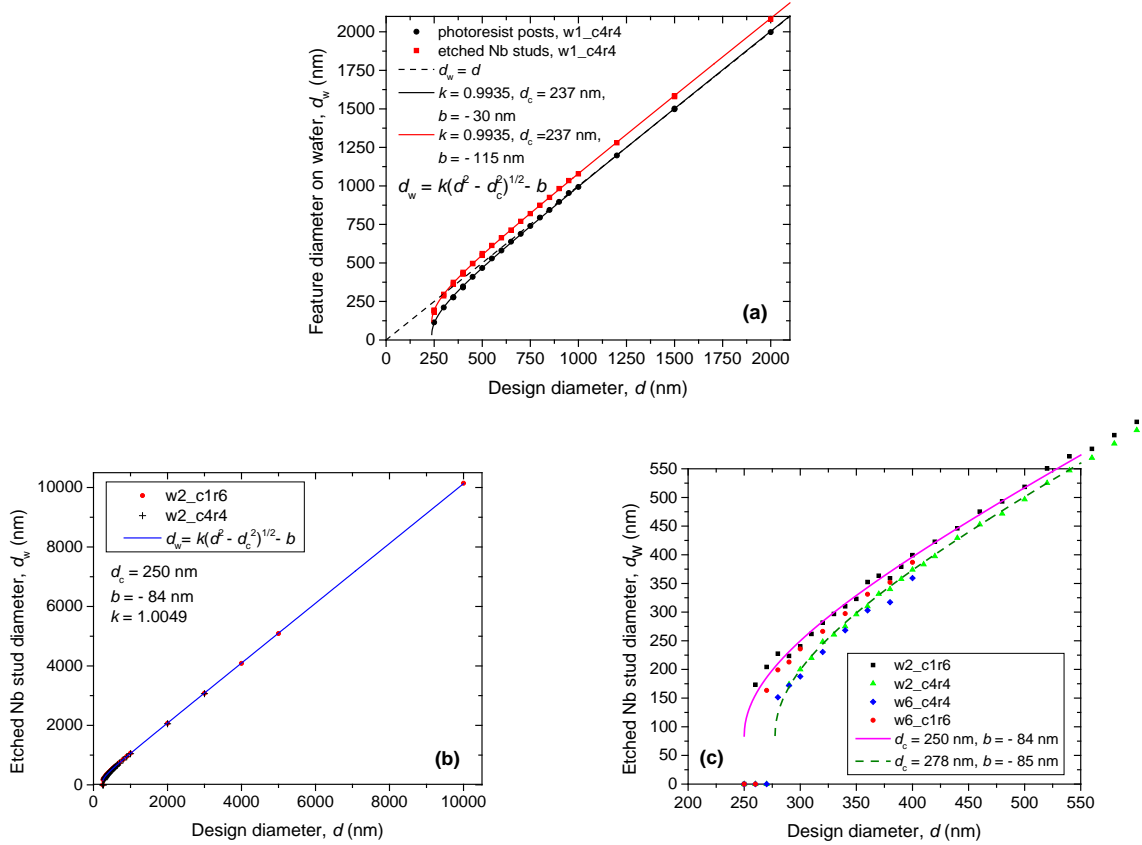

Figure 7. The mean diameter of features used for defining Nb stud-vias as a function of design diameter: (a) diameter of photoresist posts and the bottom diameter of the etched Nb studs in the center (c4r4) of wafer 1; (b) diameter of etched studs on wafer 2 in the wide range of design diameters; (c) diameter of etched Nb studs, zoom in on the range of sizes relevant for via-stud definition (≤ 600 nm) on two wafers (central die marked c4r4 and left upper die marked c1r6). Solid lines are fits to equation (1), showing the lithography cut-off of about 250 nm and a small negative process bias (the obtained features are larger than the designed).

We have found that the size of the obtained photoresist image of the opaque disc shapes on the mask can be well described by the relationship:

$$d_w = k(d^2 - d_c^2)^{1/2} - b, \quad d > d_c \qquad (1)$$

and $d = 0$ for $d \leq d_c$, where $d_c$ is a photolighography cut-off size that depends on the exposure dose, focus, and other lithography parameters, and $b$ is the process bias that can be positive or negative depending on the process conditions. Parameter $k$ is the scaling factor characterizing the accuracy of the projection optics reduction coefficient and SEM magnification, ideally $k = 1$. Based on multiple measurements on many wafers, we have found that for the stable and optimized process $d_c$ is always about 250 nm, about the nominal resolution limit of the stepper set by the light diffraction, $k = 1\pm0.7\%$, and $b \approx 0$. Far from the resolution cut-off, the relationship between $d_w$ and $d$ is perfectly linear over a very wide range of diameters.

Relationship (1) has some implications for the design and fabrication of JJs and other small objects in circuits. It is usually assumed that the size of the projected objects is given simply by $d_w = d - b$. For circular objects this leads to the concept of a "missing" diameter (radius). It is often used in the description of the size dependence of such parameters as JJ critical currents and normal state conductance that are proportional to the actual area, e.g., $I_c = j_c(\pi/4)(d - b)^2$ [17],[18], where $j_c$ is the critical current density. Our measurements indicate that this concept is not adequate for the description of small features, especially near the resolution limit. At $b = 0$, equation (1) rather supports the concept of a "missing" area, also often used in circuit design, $A_w = A_d - A_c$, where $A_w$ is the object's area on the wafer, $A_d$ – the design area, and $A_c$ – a cut-off ("missing") area. The origin of the lithography cut-off is the diffraction of light. Decreasing the diameter of an opaque disc leads to a blurring of its image on the photoresist and increases the light intensity in the area of the geometrical shade. At a fixed exposure dose, $d_c$ corresponds to the size at which the diffracted intensity produces the critical exposure of the photoresist in the shade area when it will be completely developed off. Therefore, if a printing of discs with the same diameter or in the narrow range of diameters is only required, the average exposure dose can be optimized to reduce $d_c$ and allow for resolving the smaller objects, although changing the linearity of relationship between $d_w$ and $d$ at other sizes. On the other hand, if printing a range of sizes near the resolution limit is required, a proper biasing of the features on the reticle needs to be done according to equation 1 to produce a linear scaling on the wafer.

*3.2 Etching*

Nb etching was done using the high density plasma etching chamber of an Applied Materials Centura system with end-point detection, stopping on the Al etch-stop layer. After stripping the photoresist, the diameter of the Nb studs at their bottom, $d_w$, was measured. The results are shown in figure 7. As can be seen, after etching the bottom diameter of studs follows the same dependence on the design (drawn) diameter [equation (1)] as the photoresist mask. Only the process bias $b$ becomes slightly more negative (by about 85 nm) after etching, showing that the features after etching are slightly larger than the photoresist mask. The total process bias is thus the same for all drawn diameters, $b = b_{photo} + b_{etch}$, and is about −100 nm, see figure 7(a) – figure 7(c). The diameter of studs at their tops was measured to be less than at the bottom by between 50 nm and 70 nm. This gives the side wall angle in the range from 80° to 84°. Therefore, in the following discussion we will consider cylindrical studs with a diameter equal to the mean of the bottom and top diameters, $d_{av} \approx d_w - 35$ (nm). Hence, the effective diameter $d_{av}$ is given by the same Eq. (1) with a modified total process bias (about −80 nm for studs in figure 7(a), and about −50 nm for studs in figures 7(b) and 7(c)).

Next, the photolithography and etching of the bottom electrode was done to define the bottom wiring layer. We used a few different design rules for the surround, $s$, of studs by the bottom and top Nb (see below). The minimum surround used was 100 nm, close to the typical overlay accuracy (mean+ three sigma) of our photolithography tool. For simplicity we used the same set of reticles that we normally use for the fabrication of Josephson junctions. So the surround and via placement were not specially optimized to achieve the minimum possible stud-via area. This is planned to be done in the future.

Etching of the bottom wire was done in a Cl-based chemistry to break through the Al etch-stop layer first. SEM images of the obtained structures are shown in figure 8. After stripping the photoresist, a $SiO_2$ film was deposited using plasma-enhanced chemical vapor deposition (PECVD) from $SiH_4/N_2O/Ar$ mixture for subsequent planarization by CMP. The thickness of the film was about twice the thickness of the Nb/Al/Nb trilayer. CMP was done using an Applied Materials Mirra polisher to planarize and remove oxide up to the level of the tops of the Nb studs, figure 4(c). The thickness of the remaining oxide was controlled by an ellipsometer.

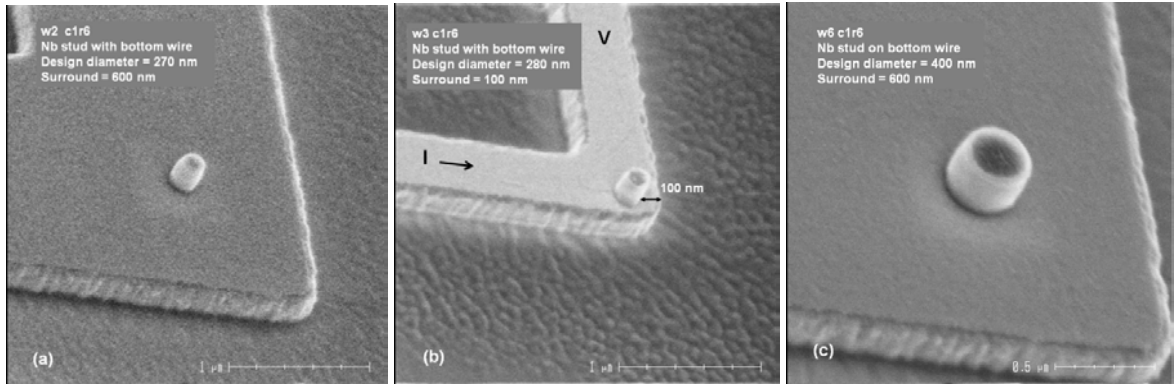

Figure 8. Etched Nb studs with various diameter on top of bottom Nb wire: (a) 270-nm design diameter, 600-nm surround; (b) 280-nm design diameter, 100-nm surround; (c) 400-nm design diameter, 600-nm surround. The bottom wire has an L-shape that with the corresponding L-shape in the top wire will form a cross-bridge Kelvin resistor (CBKR) configuration for electric measurements of studs. One arm is to apply current along the wire and up through the stud to the top wire (not shown), and another one is to measure voltage (with respect to the top wire).

Finally, the top wiring layer of Nb was deposited and patterned similarly to the bottom wiring and with the same overlap. The bottom and top wiring layers connected each stud to a cross-bridge Kelvin resistor (CBKR) geometry for 4-point measurements of room temperature resistance and critical current at LHe temperature. The SEM picture of the completed structure is shown in figure 9. Here one can clearly see the L-shape of the top Nb wire and a blurry image of the L-shape of the bottom wire which is seen through 250-nm interlayer dielectric. The outline of the 350-nm stud can be also seen as it slightly protrudes up after polishing, creating a barely visible image in the top wire.

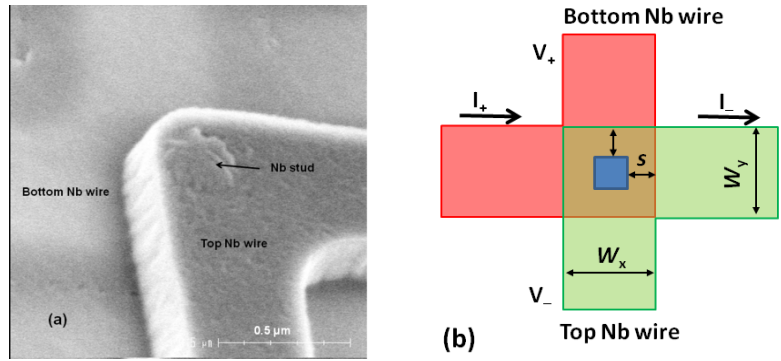

Figure 9. (a) SEM image of the completed CBKR structure with 350-nm stud between two L-shaped wires. Clearly visible is the top Nb wire, lighter blurry image is the bottom wire as seen through $SiO_2$ dielectric. The outline of the stud can be also seen due to a difference in niobium grain structure. (b) A sketch of CBKR geometry with a rectangular contact used in the analytical model in [19]. The actual shape of the contact between the studs and the top and bottom wires is circular as in figures 8 and 9(a).

## 4. Electrical test results and discussion

*4.1 Resistance in the normal state*

Electric measurements of the Kelvin resistance of studs $R_K = (V_+ - V_-)/I$, where $I$ is the applied current as shown in figure 9, were done using a semi-automated wafer prober. Because of a very large number of test structures we fully probed a limited set of 9 dies out of 49 across 200-mm wafers. The measurements indicated good uniformity of the fabricated studs across the wafers. The typical results are presented in figure 10.

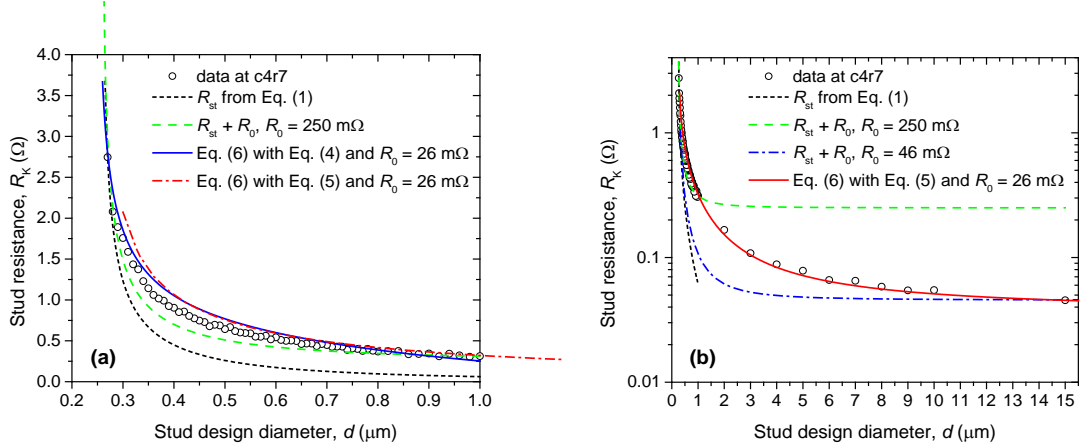

Figure 10. Electrical properties of Nb studs: (a) room temperature Kelvin resistance of the stud structure shown in Fig. 9 as a function of design diameter of studs in the range relevant to the stud-via process, $d < 1$ µm; (b) full range of the studied diameters. Dotted lines show the expected stud resistance $R_{st}$ from equation (2) at $t = 0.25$ µm and $R_s = 0.86$ Ω/sq. Dash lines shows the effect of a constant parasitic resistance $R_0$ in series with stud resistance $R_{st}$ for two values of $R_0$. Solid lines are the fits to equation (6) with $R_0$ as the only fitting parameter (see text).

Since we are interested in the scaling of the stud conductance with area, in figure 10 we plotted $R_K$ versus the design diameter of studs. In the simplest (ideal) case (when all parasitics can be neglected) the measured CBKR resistance is the resistance of the stud given by

$$R_{st} = 4R_s t^2/(\pi d_{av}^2), \qquad (2)$$

where $t$ is the stud height (250 nm), $R_s$ is the sheet resistance of Nb film from which the studs were etched, and $d_{av}$ is the average diameter of the stud on wafer that is related to design diameter $d$ as was discussed in section 3. We assume that the resistivity, $\rho$, of Nb in studs after etching is nearly the same as the resistivity of the initial 200-nm Nb film ($R_s \approx 1.07$ Ω/sq) in the trilayer, giving $R_s = 0.86$ Ω/sq for the studs. Equation (2) at $t = 250$ nm and $R_s = 0.86$ Ω/sq, and with $d_{av}$ given by equation (1) at $d_c = 0.255$ µm and $b = -85$ nm is shown in figure 10 by a dotted line. It describes the measured dependence only at very small diameters of studs [figure 10(a)] when their resistance is high and dominates the total resistance measured by the CBKR structure.

At large diameters, the measured resistance is much higher than that given by equation (2) and clearly has a tendency to saturation as stud diameter increases. This could be explained if there exists a small constant parasitic resistance $R_0$ in series with the stud resistance, so the total measured resistance is $R_{st} + R_0$, which approaches $R_0$ with increasing the stud diameter. Although, the origin of this parasitic is not exactly clear, we noted its presence, with $R_0$ in the range from ~ 20 mΩ to ~ 50 mΩ, in all of our room-temperature measurements of studs and tunnel junctions using the described CBKR geometry. The effect of this parasitic series resistance is shown in figure 10(a) by dash lines for two values of $R_0$. Although a better agreement could be reached in a narrow range of diameters [e.g., at $R_0$ = 250 mΩ, figure 10(a)], this is clearly not sufficient to describe the measured dependence in a wider range of diameters of studs.

The CBKR geometry shown in figure 9 is frequently used for measurements of contact resistance between two materials or the resistance of various barriers between two materials. There is a great deal of literature devoted to its analysis, both analytical and numerical (see, e.g., [19] and references therein), particularly in cases when the barrier can be considered two-dimensional, e.g., an opening in a non-conducting dielectric of very small thickness separating the top and bottom films as shown in figure 9(b). In the superconducting state, when there is no voltage drop associated with current redistribution (crowding) in the wires near the contact, it measures exactly the resistance of the barrier.

In all other cases, the measured Kelvin resistance $R_K$ includes parasitics associated with two-dimensional and three-dimensional current crowding effects. The analytic expression for the 2D parasitic resistance associated with current crowding in both films due to the finite surround of a rectangular contact was given in [19] as

$$R_{geom} = (8/3)R_{wire}[s^2/W_x W_y][1+0.5s/(W_x-s)], \qquad (3)$$

where $W_x$ and $W_y$ are the widths of the voltage and current electrodes, respectively, $s$ is the surround shown in figure 9(b) and $R_{wire}$ is the sheet resistance of the top and bottom wire films assumed to be identical.

In order to make a comparison with equation (3) we replace the circular contact with diameter $d_{av}$ by a square contact with the same area. In our stud-via test structures the width of the wires was constant $W_x = W_y = W = 1.7$ µm for $d \leq 0.8$ µm, giving an effective surround $s = [W-(\sqrt{\pi}/2)d_{av}]/2$ and

$$R_{geom} = (2/3)R_{wire}[1-(\sqrt{\pi}/2)d_{av}/W]^2\{1+0.5[W-(\sqrt{\pi}/2)d_{av}]/[W+(\sqrt{\pi}/2)d_{av}]\}. \qquad (4)$$

At larger diameters, $d > 0.8$ µm, we kept a constant design surround $s_0 = 0.425$ µm, so $W = d+2s_0$ and $s = [d+2s_0-(\sqrt{\pi}/2)d_{av}]/2$, resulting in

$$R_{geom} = (2/3)R_{wire}\{[d+2s_0-(\sqrt{\pi}/2)d_{av}]/(d+2s_0)\}^2\{1+[d+2s_0-(\sqrt{\pi}/2)d_{av}]/[d+2s_0+(\sqrt{\pi}/2)d_{av}]\}, \qquad (5)$$

We fitted the measured resistance $R_K$ to the sum of all three contributions

$$R_K = R_{st} + R_{geom} + R_0, \qquad (6)$$

with $R_{geom}$ given by equation (4) at $d \leq 0.8$ µm [figure 10(a)] and equation (5) at $d > 0.8$ µm [figure 10(b)], and treating $R_0$ as the only fitting parameter. We used the measured value of the sheet resistance for the top and bottom wires $R_{wire} = 1.07$ Ω/sq and the same value of Nb resistivity for the studs, giving $R_s = 0.86$ Ω/sq for studs. The obtained fits to equation (6) are shown in figure 10 by solid lines for the two ranges of stud diameters. As can be seen, the overall description of the data is very good in a very wide range of stud diameters and with only one fitting parameter.

*4.2 Superconducting critical current*

The critical current, $I_c$, of the fabricated studs (figure 11) was measured in liquid helium by taking $I$-$V$ characteristics. The transition into the resistive state at $I_c$ is very sharp and the return into the superconducting state occurs at a much lower current indicating significant thermal hysteresis.

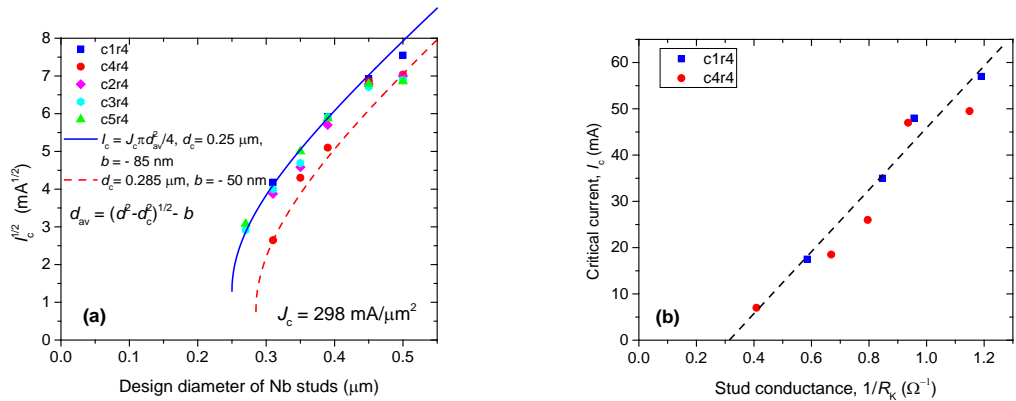

Figure 11. Critical current of the fabricated Nb stud-vias as a function of the design diameter for several dies at different locations across the wafer (a). The measurements were restricted to the range of stud-via diameters expected to be used in superconducting integrated circuits, $d \leq 0.5$ µm. Fits to the dependence $I_c = J_c^*(\pi/4)d_{av}^2$ with $J_c = 0.3$ A/µm$^2$ are shown for the two values of the lithography cut-off parameter and process bias: $d_0 = 250$ nm $b = -85$ nm, and $d_0 = 290$ nm $b = -50$ nm. (b) The critical currents of the fabricated studs as a function of their conductance ($1/R_K$) at room temperature.

From figure 11 we can see that $I_c$ scales properly with the actual area of the studs, $I_c = J_c *$Area. The current density $J_c$ was found to be $3\cdot10^7$ A/cm$^2$ (0.3 A/µm$^2$) and only about a factor of 2.5 lower than the Ginzburg-Landau depairing critical current density $J_c^{GL} = (2/3)^{3/2}H_c/(\mu_0\lambda)$ in our films, where $\mu_0 = 4\pi\cdot10^{-7}$ H/m, $H_c$ - the thermodynamic critical magnetic field, $\lambda$ – the magnetic field penetration depth [20]. Indeed, for Nb at 4.2 K, $H_c \approx 1450$ G [21] and $\lambda$ is in the range from 80 nm to 90 nm for our films. This gives $J_c^{GL}$ in the range from $7.4\times10^7$ A/cm$^2$ to $8.3\times10^7$ A/cm$^2$, see also [22]. In the actual structure, the critical current density should be somewhat lower than $J_c^{GL}$ of the film due to the presence of an 8-nm Al etch-stop layer at the interface between the Nb stud and the bottom Nb wire, making the structure an *SS'S* junction which properties depend on the parameters of the *S* and *S'* layers and the interface resistance between them [23].

On the other hand, in the studied range, the diameter of the studs is larger than $4\xi$, where $\xi$ is the coherence length that is about 20 nm in our films. Hence, instead of the depairing, the stud critical current can be caused by the entry and motion of Abrikosov vortices when the current-induced magnetic field at the stud surface reaches $H_{c1}$. This gives an estimate for the critical current density $j_c = H_{c1}/(\mu_0\lambda)$. Using $H_{c1} = H_c\ln(\kappa+0.08)/(\sqrt{2}\kappa)$ [24] with $\kappa = \lambda/\xi = 4$ for our films, we get $H_{c1} \approx 0.25H_c$ and $j_c \approx 0.36$ A/µm$^2$ in a perfect agreement with the critical current density observed in the fabricated Nb studs.

Independently of the actual critical current mechanism, the observed critical currents are more than sufficient for the use of stud-vias in superconducting integrated circuits for interlayer connections. At $d = 0.5$ µm the critical currents of stud-vias exceed those observed in etched contact holes of the same diameter filled with deposited Nb metal.

## 5. Conclusions

We have developed and demonstrated a novel process for making deep sub-µm superconducting multilayer interconnects for use in VLSI and ULSI superconductive digital circuits. These interconnects are formed using Nb/Al/Nb trilayer wiring layers by etching Nb studs in the top layer and Nb wires in the bottom layer of the trilayers (a dual-etch process) with subsequent planarization of the formed interconnects by dielectric CMP (DEAP process). The purpose of this process development is to replace the currently used etched-contact holes filled with sputtered Nb that are too big for VLSI SFQ circuits.

Nb stud-vias with diameters as small as 150 nm have been fabricated by the developed process. For design stud diameter of 280 nm and above, the yield of the fabricated Nb stud-vias was 100% on the test structures available. Critical currents of the obtained Nb stud-vias approach the maximum possible superconducting currents for Nb – Ginzburg-Landau depairing current − and are certainly sufficient for their use as interconnects in multilayered VLSI circuits.

We presented a detailed characterization of the stud photolithography process near the resolution limit of 248-nm photolithography as well as electric characterization at room temperature of the studs in a CBKR configuration, emphasizing the importance of parasitics related to current crowding around the studs. The performed analysis is also applicable to the fabrication of deep sub-µm Josephson junctions and to the characterization of their room temperature resistance for purposes of the process control and monitoring, especially in cases of junctions with high Josephson critical current densities when their tunnel resistance is low and the measured Kelvin resistance $R_K$ is dominated by parasitics.

We proposed two versions of the stud-via process and presented a practical realization of one of them. The only other feasible alternative to the described processes would be a dual-damascene-type process with vias and trenches etched in interlayer dielectric and filled by an advanced PVD (or CVD) process for Nb, and followed by CMP of Nb. It remains to be proven if such a fill process and CMP process can be developed and produce reliable and superconducting interconnects.

**Acknowledgement**
The authors would like to thank Dr. Marc Manheimer and Dr. Scott Holmes for their interest and support of this work.